\documentclass[11pt]{article}

\usepackage[final]{acl}

\usepackage{times}
\usepackage{latexsym}

\usepackage[T1]{fontenc}

\usepackage[utf8]{inputenc}

\usepackage{microtype}

\usepackage{inconsolata}

\usepackage{graphicx}
\graphicspath{{figures/}}
\usepackage{amsmath}
\usepackage{amssymb}
\usepackage{adjustbox}

\title{IF-GEO: Conflict-Aware Instruction Fusion for Multi-Query Generative Engine Optimization}

\author{
  Heyang Zhou\textsuperscript{1}, 
  JiaJia Chen\textsuperscript{2}, 
  Xiaolu Chen\textsuperscript{1}, 
  Jie Bao\textsuperscript{1}, 
  \textbf{Zhen Chen}\textsuperscript{1}\thanks{~~Corresponding author.}, 
  \textbf{Yong Liao}\textsuperscript{1} \\
  \textsuperscript{1}School of Cyber Science and Technology, University of Science and Technology of China \\
  \textsuperscript{2}Institute of Dataspace, Hefei Comprehensive National Science Center, Hefei, China \\
  \texttt{heyangzhou@mail.ustc.edu.cn}
}

\usepackage{booktabs}  
\usepackage{multirow}  
\usepackage{amsmath}

\usepackage[skins,breakable]{tcolorbox}
\usepackage{enumitem}

\usepackage{tabularx}

\tcbset{
  ifgeoGray/.style={
    enhanced,
    breakable,
    boxrule=0.4pt,
    arc=2pt,
    left=6pt,right=6pt,top=6pt,bottom=6pt,
    colback=black!2,
    colframe=black!25,
    before skip=4pt,
    after skip=8pt,
    fontupper=\footnotesize,
    before upper=\raggedright
  }
}

\begin{document}
\maketitle

\begin{abstract}
As Generative Engines revolutionize information retrieval by synthesizing direct answers from retrieved sources, ensuring source visibility becomes a significant challenge. Improving it through targeted content revisions is a practical strategy termed Generative Engine Optimization (GEO). However, optimizing a document for diverse queries presents a constrained optimization challenge where heterogeneous queries often impose conflicting and competing revision requirements under a limited content budget. To address this challenge, we propose IF-GEO, a \emph{``diverge-then-converge''} framework comprising two phases: (i) mining distinct optimization preferences from representative latent queries; (ii) synthesizing a \emph{Global Revision Blueprint} for guided editing by coordinating preferences via conflict-aware instruction fusion. To explicitly quantify IF-GEO's objective of cross-query stability, we introduce risk-aware stability metrics. Experiments on multi-query benchmarks demonstrate that IF-GEO achieves substantial performance gains while maintaining robustness across diverse retrieval scenarios.
\end{abstract}

\section{Introduction}
\label{sec:intro}

\begin{figure}[t]
  \centering
  \includegraphics[width=\columnwidth]{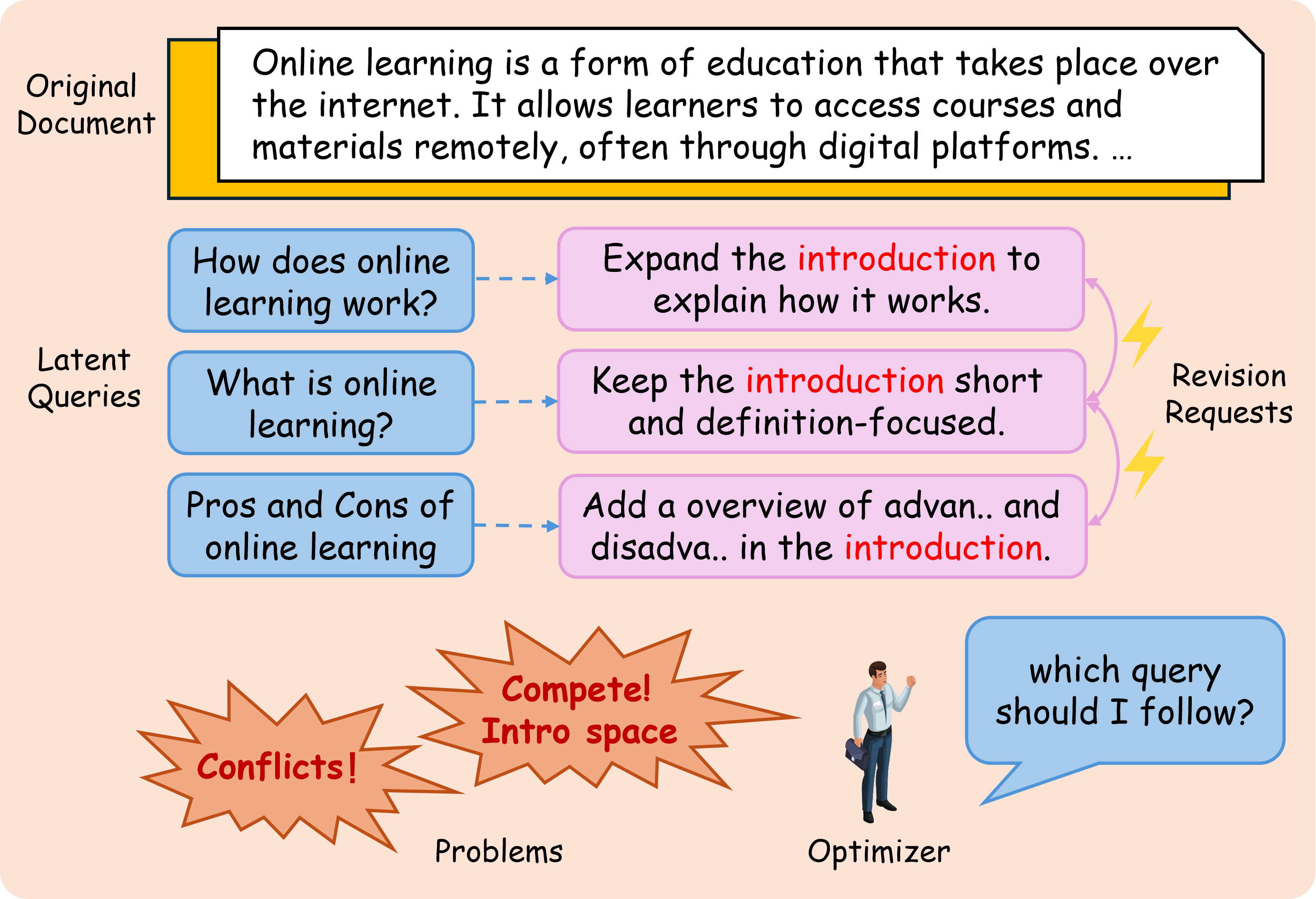}
  \caption{Challenges of GEO. Revision requests of different queries can be \emph{conflicting} and \emph{competitive} under a limited content budget. GEO have no idea which query to follow.}
  \label{fig:intro}
\end{figure}
The evolution from traditional Search Engines (SEs) to Generative Search Engines (GSEs) represents a significant shift in information retrieval~\citep{brin1998anatomy,aggarwal2024geo}. Unlike SEs, which present ranked lists of hyperlinks, GSEs adopt a retrieve-then-generate paradigm~\citep{lewis2020rag,karpukhin2020dpr}. By employing Large Language Models (LLMs) to synthesize direct answers from retrieved documents, GSEs provide users with synthesized information rather than a list of sources.

However, ensuring source visibility presents a significant challenge for content providers. Since visibility depends on whether a document is selected and cited by the model rather than its ranking position, traditional optimization techniques are less effective. To address this, Generative Engine Optimization (GEO) has been proposed as a prominent strategy to improve content visibility in generated responses through targeted document revisions~\citep{aggarwal2024geo}.

Existing GEO methods typically utilize static heuristic rules~\citep{aggarwal2024geo} or optimize content based on aggregated engine preferences inferred from ranking signals~\citep{wu2025preferences, chen2025dominate}. 
While recent work like RAID~\citep{chen2025intentdriven} advances the field by modeling latent retrieval intents, it prioritizes a single aggregated intent trajectory. 
These approaches treat the serving of diverse heterogeneous queries as a one-dimensional optimization problem.
However, in practice, a single document needs to serve diverse user queries simultaneously~\citep{broder2002taxonomy,clarke2008intent,wang2009mining}. This creates a constrained optimization problem, where heterogeneous queries often impose conflicting requirements on the document's limited content as illustrated in Figure~\ref{fig:intro}~\citep{marler2004mooSurvey,rose2004goals}. Current methods lack the mechanisms to coordinate these competing preferences.
Consequently, optimizing for one query often results in diminished or negative gains for others, creating competitive imbalances that we empirically analyze in Appendix~\ref{sec:appendix_competition}, leading to inconsistent revisions and performance variance across the query set.

To address this challenge, we propose \textbf{IF-GEO}, a \emph{``diverge-then-converge''} framework comprising two phases: (i) mining representative latent queries and formulating their distinct optimization preferences as structured edit requests; and (ii) synthesizing a unified \emph{Global Revision Blueprint} for guided editing by coordinating preferences via conflict-aware instruction fusion. 
Crucially, we formally integrate risk-aware stability metrics into the IF-GEO optimization objective to explicitly quantify cross-query stability. By introducing Worst-Case Performance (WCP), Win-Tie Rate (WTR), and Downside Risk (DR), we establish a robust standard that addresses the limitations of mean-variance evaluations---which often mask tail degradations and conflate beneficial upside with harmful downside volatility.

Our contributions are summarized as follows:
\begin{itemize}
    \item We propose \textbf{IF-GEO}, a \emph{``diverge-then-converge''} framework. It predicts latent queries to formulate specific edit requests and employs conflict-aware instruction fusion to synthesize a unified \emph{Global Revision Blueprint} for coherent document revision.
    \item We introduce an evaluation methodology utilizing risk-aware stability metrics—specifically Worst-Case Performance, Downside Risk, and Win-Tie Rate—to explicitly quantify the safety and robustness of optimization across diverse queries.
    \item Empirical results on multi-query benchmarks demonstrate that IF-GEO achieves substantial improvements in overall visibility while effectively mitigating performance variance, ensuring robust stability across heterogeneous retrieval scenarios.
\end{itemize}

\section{Related Works}
\label{sec:related}

\subsection{Generative Search Engines}
Generative Search Engines (GSEs) increasingly follow a \textit{retrieve--then--generate} paradigm, producing answer-centric responses with explicit source attributions. Technically, many such systems build on retrieval-augmented generation (RAG) \citep{lewis2020rag} and retrieval-augmented pretraining/memory architectures such as REALM \citep{guu2020realm} and RETRO \citep{borgeaud2022retrieval}. In practice, these pipelines often pair dense neural retrievers (e.g., DPR) \citep{karpukhin2020dpr} with multi-passage evidence fusion generators (e.g., FiD) \citep{izacard2021fid}. Unlike traditional search engines that prioritize document ranking based on keyword matching\citep{gleason2023gatekeeper}, GSEs leverage the semantic reasoning capabilities of LLMs to synthesize information from multiple disparate sources \citep{li2024genirSurvey}. To improve verifiability and citation consistency, prior work has introduced browsing/citation constraints and dedicated benchmarks for attribution quality (e.g., WebGPT \citep{nakano2021webgpt}, GopherCite \citep{menick2022gophercite}, and CITE \citep{gao2023cite}). These architectural shifts fundamentally alter the optimization landscape, motivating downstream objectives that emphasize not just retrieval rank, but visibility and attribution within the generated narrative.

\subsection{Generative Engine Optimization}
Generative Engine Optimization (GEO) aims to improve visibility within generative answers through content modifications. Existing work broadly falls into three lines. 
(i) Single-objective and heuristic optimization. Early work formalized GEO with heuristic strategies \citep{aggarwal2024geo}, while recent works explored specific interventions like caption injection \citep{chen2025caption} or transformer-based rewriting \citep{luttgenau2025beyondseo}. These approaches apply preset or single-target editing rules, lacking adaptation to diverse query conflicts.
(ii) Feedback-driven Optimization. This line treats engines as black boxes, optimizing content by inferring implicit preferences from ranking signals \citep{wu2025preferences} or employing iterative feedback loops \citep{bagga2025egeo, chen2025dominate}. However, they often overfit to specific engine behaviors and ignore cross-query stability.
(iii) Intent-based optimization. This line considers optimization over a document’s latent queries.
RAID leverages multi-role reflection to generalize latent search intents\citep{chen2025intentdriven}, but optimizes around a single aggregated intent. Overall, prior methods treat optimization targets in isolation or aggregation. They lack mechanisms to coordinate diverse intents, failing to balance competitive trade-offs under a limited content budget, which ultimately hinders robust and stable optimization.

\section{Problem Definition}
\label{sec:problem}
\subsection{Setting and Notation}
\label{sec:setting}

Given a query expression $q$, a generative engine (GE) retrieves a set of candidate documents
$\mathcal{C}_q = \mathrm{Retrieval}(q) = \{ D_1, \ldots, D_N \}$
and generates a response text
$r_q = \mathrm{Generate}(q, \mathcal{C}_q)$,
which includes citations to documents in $\mathcal{C}_q$.
We define an answer-induced visibility score
$v(D_i, r_q)$
to quantify how visible a document $D_i \in \mathcal{C}_q$ is within the response $r_q$,
based only on the response text and its citation/attribution signals.
Since $r_q$ is conditioned on $q$, we use the shorthand $v(D, q)$
to denote the visibility of a document $D$ under query $q$.
Importantly, $v(\cdot)$ is a \emph{plug-in} metric and can be instantiated with different
operational definitions of visibility.

\subsection{Optimization Objective}
\label{sec:multi_intent_obj}

Given a document $D$, we consider a target set of queries
$Q(D)={q_i}_{i=1}^{m}$ that represent the queries we aim to optimize for.
For each $q_i\in Q(D)$, the visibility of $D$ under $q_i$ is measured by $v(D,q_i)$.

A GEO method $\mathcal{M}$ takes $D$ as input and produces an optimized document $D'=\mathcal{M}(D)$.
For each query $q_i\in Q(D)$, we define the per-query visibility gain as:
\begin{equation}
\Delta v_i \ =\ v(D',q_i)-v(D,q_i).
\end{equation}
We summarize the optimization effect over the query set $Q(D)$ by applying a plug-in aggregation functional $\mathcal{A}(\cdot)$ to the gain vector:
\begin{equation}
\Delta \ =\ \mathcal{A}\big(\{\,\Delta v_i\,\}_{i=1}^{m}\big).
\end{equation}
The choice of $\mathcal{A}$ is application-dependent and can be instantiated with different operational definitions (e.g., mean for average improvement).
GEO aims to find a method $\mathcal{M}$ that optimizes $\Delta$ under the chosen definition of $\mathcal{A}$.

\section{Method}
\label{sec:method}

\begin{figure*}[t]
  \centering
  \includegraphics[width=\textwidth]{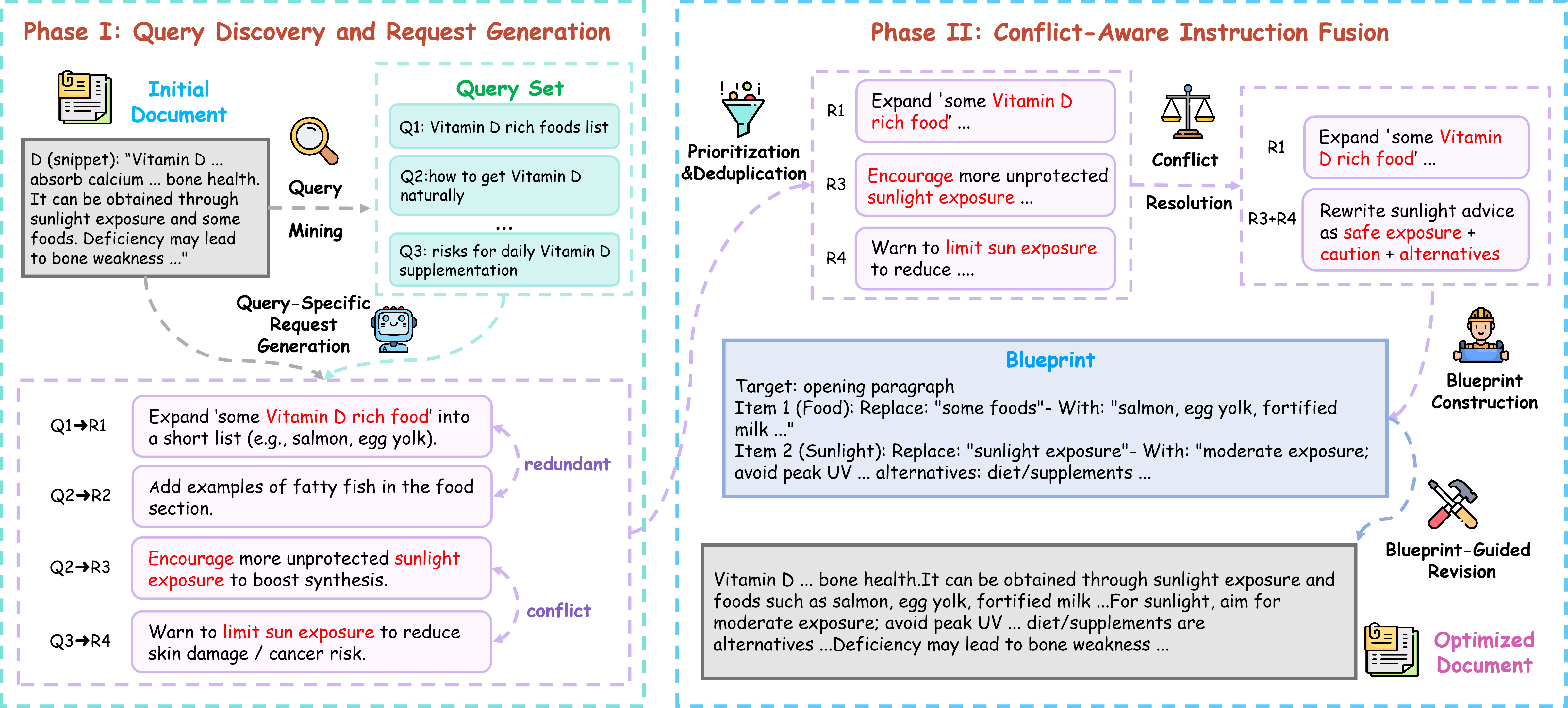}
  \caption{Overview of the IF-GEO methodology. 
  IF-GEO follows a ``diverge-then-converge'' paradigm. 
  In Phase I, the system mines a representative query set and elicits query-specific editing requests, which may be redundant or conflicting. 
  In Phase II, IF-GEO performs conflict-aware instruction fusion to synthesize a unified global revision blueprint, which guides controlled editing to produce an optimized document with stable visibility across queries.}
  \label{fig:method}
\end{figure*}

\subsection{Overview: IF-GEO}
\label{sec:overview}

We propose \textbf{IF-GEO}, a two-stage framework for GEO under generative engines.
IF-GEO follows a \emph{``diverge-then-converge''} workflow. In the \emph{diverge} phase, it discover representative latent queries and elicit their distinct optimization preferences as structured edit requests. The \emph{converge} phase then coordinates and balances these competing preferences through conflict-aware instruction fusion, synthesizing them into a unified \emph{global revision blueprint} that provides a comprehensive optimization direction for guided editing.
Beyond maximizing average visibility, IF-GEO explicitly optimizes for stability metrics (e.g., WCP, DR) to minimize cross-query downside risks, ensuring robust performance where standard mean-based approaches often fail.

Figure~\ref{fig:method} overviews the pipeline.
All steps of IF-GEO are implemented via LLM calls, where the model acts as a constrained executor that outputs structured intermediate artifacts rather than performing free-form rewriting. Full prompt specifications and output schemas are provided in Appendix B.

We next detail the two phases of this workflow: Query Discovery and Request Generation as the diverge phase, followed by Instruction Fusion as the converge phase.

\subsection{Query Discovery and Request Generation}
\label{sec:elicitation}

\subsubsection{Query Mining}
As the first step in the \emph{diverge} phase, we conduct \textbf{query mining} to expand the document into a representative query set. We treat this process as \textbf{reverse retrieval}: given $D$, we guide the LLM to act as a search analyst and discover diverse queries for which the document should be a relevant result.
To ensure the set is representative and avoid redundant tuning, the system explicitly instructs the model to target different aspects of the document's main theme while strictly prohibiting simple paraphrases. This step outputs a weighted query set $Q(D)$:
\begin{equation}
Q(D) = \{(q_i, w_i)\}_{i=1}^{m}
\end{equation}
where each query $q_i$ serves as a specific context for generating subsequent edit requests. The associated weight $w_i$ is a scalar score (0--100) estimating query popularity. Drawing on the scoring capability demonstrated in G-EVAL \citep{liu2023geval}, we leverage the model's parametric knowledge to assign this priority, which guides the later fusion phase.

\subsubsection{Query-specific Request Generation}
\label{sec:request_generation}

Given the weighted query set $Q(D)$, the core objective of this step is to elicit the distinct optimization preferences of each query regarding the document. IF-GEO operationalizes these abstract preferences into concrete, actionable \emph{edit requests} (instruction candidates). By analyzing each query $q_i$ in isolation, the model explicitly diagnoses specific content gaps and proposes targeted revisions to address them.

Formally, for the $j$-th request derived from query $q_i$, we define a structured request tuple:
\begin{equation}
r_{i,j} = \langle e_{i,j}, u_{i,j}, s_{i,j} \rangle
\end{equation}
where $e_{i,j}$ is the excerpt-based anchor locating the target text, $u_{i,j}$ is the specific revision suggestion, and $s_{i,j}$ is a quantitative \emph{necessity score} (0--100). Adopting the scalar evaluation methodology validated in \citep{liu2023geval}, we employ this score to explicitly measure the criticality of the fix for satisfying $q_i$, distinguishing essential edits from marginal improvements. Aggregating these outputs yields a diverged request pool $\mathcal{R} = \{r_{i,j}\}$, which serves as the raw material for the subsequent conflict-aware fusion.

\subsection{Conflict-Aware Instruction Fusion}
\label{sec:synthesis}

\subsubsection{Prioritization and Deduplication}
\label{sec:prioritization}
The converge phase begins by consolidating the diverged request pool $\mathcal{R}$ into a compact, high-signal candidate set. First, to identify requests with high cross-query importance, we compute a \emph{global priority score} for each item:
\begin{equation}
g_{i,j} = w_i \cdot s_{i,j}
\end{equation}
where $w_i$ is the query weight and $s_{i,j}$ is the necessity score. We filter out low-priority requests falling below a threshold $\tau$ to control perturbation.
Next, we perform semantic deduplication to remove redundancy. Requests targeting overlapping anchors with similar revision goals are merged into a single meta-request. These merged items inherit the highest necessity score from their constituents and are standardized with concise topic tags. This step reduces the raw pool into a smaller set of well-scoped candidates, ready for conflict resolution.

\subsubsection{Conflict Resolution}
\label{sec:conflict}

Following deduplication, the system addresses mutually exclusive requests targeting the same content. IF-GEO employs a \emph{priority-based resolution strategy}. Instead of relying on a rigid numerical threshold, we delegate the decision to the model, instructing it to evaluate the priority scores ($g$) in context to determine the appropriate action:

\begin{itemize}[leftmargin=*]
    \item \textbf{Selection:} When the model perceives a significant priority gap, the higher-scoring instruction is strictly retained, and the conflicting one is discarded.
    \item \textbf{Synthesis:} When the model determines that priority scores are comparable ($g_i \approx g_j$), it generates a new compromise instruction. This instruction balances the valid needs of both queries rather than satisfying only one.
\end{itemize}

This approach leverages the model's semantic reasoning capabilities to handle edge cases more flexibly than hard-coded rules.

\subsubsection{Blueprint Construction}
\label{sec:blueprint}
As the final step of the \emph{converge} phase, we consolidate the fused instructions into a \textbf{Global Revision Blueprint}. This step is crucial for coordinating the competing needs of multiple queries within a shared content budget. We map individual instructions to their corresponding document sections and aggregate them into ordered plan items.
By organizing revisions structurally rather than sequentially by query, the blueprint ensures that distinct optimization goals are \textbf{integrated coherently}. The resulting JSON blueprint serves as a strict execution contract, effectively resolving the structural conflicts between queries and preventing the inconsistent edits typical of single-pass optimization.

\subsubsection{Blueprint-Guided Revision}
\label{sec:execution}

In the final step, IF-GEO uses the Revision Blueprint to generate the optimized document $D'$. The revision model acts as a constrained editor, strictly following the blueprint instructions to modify the document section by section.
Crucially, the system explicitly instructs the model to preserve all unmentioned sections exactly as is. This strict adherence ensures that the global optimization strategy is implemented accurately without introducing free-form rewriting or unintended content drift.

\subsection{Risk-Aware Optimization Objective}
\label{sec:objective}

Standard visibility maximization often masks tail degradations and conflates beneficial upside with harmful downside volatility. To address this, we formally integrate risk-aware stability metrics into the IF-GEO objective. Beyond maximizing expected gain $\mathbb{E}[\Delta v]$, we explicitly safeguard cross-query stability through three dimensions:

\begin{itemize}[leftmargin=*]
    \item \textbf{Worst Case Performance (WCP)} establishes a safety lower bound by capturing the maximum single-query drop \citep{ben2004robust}:
    \begin{equation}
    \text{WCP} = \min_{i=1}^{m} \Delta v_i.
    \end{equation}

    \item \textbf{Downside Risk (DR)} exclusively penalizes the magnitude of negative gains to distinguish harmful failures from beneficial volatility \citep{sortino1994downside}:
    \begin{equation}
    \text{DR} = \frac{1}{m}\sum_{i=1}^{m} \big(\min(0,\Delta v_i)\big)^2.
    \end{equation}
    
    \item \textbf{Win-Tie Rate (WTR)} quantifies the coverage of non-regressive optimization, serving as a proxy for Pareto optimality without collateral damage\citep{radlinski2008interleaving}:
    \begin{equation}
    \text{WTR} = \frac{1}{m}\sum_{i=1}^{m} \mathbb{I}(\Delta v_i \ge 0).
    \end{equation}
\end{itemize}

In summary, IF-GEO seeks to maximize visibility gain subject to minimizing $\text{DR}$ and maximizing $\text{WCP}$ and $\text{WTR}$.

\section{Experimental Setup}
\label{sec:exp_setup}

\subsection{Dataset}
\label{sec:dataset}

To evaluate performance in a realistic multi-query scenario, we adopt the benchmark introduced by RAID \citep{chen2025intentdriven}, which expands the widely-used GEO-Bench \citep{aggarwal2024geo}.
In this dataset, each document $D$ is originally a top-ranked result (top-5) retrieved by Google Search for a real-world source query.
Building on this high-relevance foundation, RAID extends each source query into a cluster of five related queries.
Crucially, these queries go beyond simple rephrasing that they constitute a multifaceted inquiry into the document's topic, targeting \textbf{distinct informational dimensions} (e.g., definition, usage, pros/cons) rather than mere lexical variations.
This construction forms a query set that effectively serves as a \textbf{concrete instantiation} of the \emph{representative query set} described in our problem definition. 
Using this dataset allows us to move beyond single-query evaluation and test whether IF-GEO's optimized document maintains high visibility and stability across \textbf{diverse user queries}.

\begin{table*}[t]
  \centering
  \caption{Comparison of IF-GEO with baseline methods on objective and subjective visibility improvements. We report overall improvement (Mean) to summarize average gains. \underline{Underlined} indicates the best baseline.}
  \label{tab:main_results_updated}
  \resizebox{\textwidth}{!}{
      \begin{tabular}{l | c c c | c c c c c c c c}
        \toprule
        \multirow{2}{*}{\textbf{Method}} &
        \multicolumn{3}{c|}{\textbf{Objective Impression}} &
        \multicolumn{8}{c}{\textbf{Subjective Impression}} \\
        \cmidrule(lr){2-4} \cmidrule(l){5-12}
         & Word & Position & \textbf{Overall} & Rel. & Infl. & Unique & Div. & FollowUp & Pos. & Count & \textbf{Average} \\
        \midrule
        Tran. SEO   & 1.83 & 1.77 & 1.84 & 1.44 & 1.62 & 1.24 & 1.50 & 1.44 & 1.73 & 1.58 & 1.51 \\
        Uniq. Word  & -0.70 & -1.00 & -0.79 & -0.20 & -0.16 & -0.44 & -0.42 & -0.46 & -0.71 & -0.38 & -0.39 \\
        Simp. Expr. & 0.16 & 0.28 & 0.28 & 0.59 & 0.84 & 0.73 & 0.72 & 0.35 & 1.02 & 0.57 & 0.69 \\
        Auth. Expr. & 0.97 & 0.88 & 0.92 & 0.56 & 1.08 & 0.60 & 0.86 & 0.57 & 1.29 & 0.74 & 0.81 \\
        Flue. Expr. & 0.92 & 0.94 & 1.03 & 0.52 & 0.82 & 0.48 & 0.48 & 0.29 & 1.23 & 0.56 & 0.63 \\
        Term. Addi.  & 1.00 & 1.07 & 1.17 & 0.71 & 1.22 & 0.50 & 0.73 & 0.43 & 1.06 & 0.63 & 0.75 \\
        Cite Sources & 4.47 & 4.59 & 4.71 & 3.17 & 3.45 & 3.36 & 3.16 & 2.96 & 3.83 & 3.26 & 3.31 \\
        Quot. Addi. & 4.29 & 4.19 & 4.23 & 2.57 & 2.87 & 3.10 & 2.45 & 2.25 & 3.28 & 2.48 & 2.71 \\
        Stat. Addi. & 3.28 & 3.39 & 3.49 & 2.15 & 2.48 & 2.54 & 1.97 & 1.89 & 3.01 & 2.16 & 2.31 \\

        RAID & 1.06 & 0.78 & 0.88 & 1.44 & 1.75 & 1.40 & 1.08 & 0.84 & 1.68 & 1.34 & 1.36 \\
        Auto-GEO
        & \underline{7.80} & \underline{7.64} & \underline{7.59}
        & \underline{4.99} & \underline{5.68} & \underline{5.18} & \underline{4.74} & \underline{4.19} & \underline{6.63} & \underline{4.77} & \underline{5.30} \\

        \midrule
        \textbf{IF-GEO} &
        \textbf{11.07} & \textbf{11.15} & \textbf{11.03} &
        \textbf{5.12} & \textbf{6.16} & \textbf{5.90} & \textbf{5.29} & \textbf{5.34} & \textbf{7.32} & \textbf{5.98} & \textbf{5.87} \\
        \bottomrule
      \end{tabular}
  }
\end{table*}

\subsection{Baselines}
\label{sec:baseline}

We benchmark IF-GEO against three representative methods covering the primary paradigms in Generative Engine Optimization:

Heuristic-based: GEO \citep{aggarwal2024geo}.
Representing static, query-agnostic optimization, this framework applies nine human-designed heuristic rules. We evaluate all nine strategies, ranging from stylistic interventions (e.g., \textsc{Authoritative}) to content enrichment (e.g., \textsc{Statistics Addition}, \textsc{Cite Sources}).

Intent-based: RAID G-SEO \citep{chen2025intentdriven}.
Representing intent-driven optimization, RAID employs a ``4W'' multi-role reflection mechanism to infer and refine latent user search intents. It guides content rewriting by deepening a single, focused intent trajectory.

Preference-based: Auto-GEO \citep{wu2025preferences}.
Representing data-driven optimization, this framework automatically learns generative engine preferences from large-scale ranking data. We apply the learned, engine-specific rule sets to optimize target documents based on global utility signals.

\vspace{0.5em}
\noindent For a fair comparison, all baselines and IF-GEO are implemented using the same underlying LLM (GPT-4o-mini) and decoding configurations to isolate the impact of the optimization methodology.

\subsection{Evaluation Metrics}
\label{sec:metrics}

Following the protocol in GEO-bench \citep{aggarwal2024geo}, we evaluate both overall visibility and cross-query stability. For each query $q_i$, we compute the visibility score $v(D, q_i)$ and the optimization gain $\Delta v_i$.

\paragraph{Visibility Instantiations $v(\cdot)$.}
We instantiate $v(\cdot)$ with: (i) \textbf{Objective Impression}, the Position-Adjusted Word Count (PAWC) that weights cited text volume by citation position; and (ii) \textbf{Subjective Impression}, a qualitative LLM-as-a-judge score implemented via G-EVAL \citep{liu2023geval} and normalized to match PAWC statistics following \citep{aggarwal2024geo}.

\paragraph{Aggregation Strategies $\mathcal{A}(\cdot)$.}
To comprehensively evaluate the gain vector $\{\Delta v_i\}_{i=1}^{m}$, we instantiate the aggregation functional $\mathcal{A}(\cdot)$ (defined in \S\ref{sec:problem}) using distinct criteria. We report \textbf{Mean} and \textbf{Variance} to measure general effectiveness and volatility. Crucially, to evaluate the safety objectives formulated in \S\ref{sec:objective}, we report Worst-Case Performance (WCP), Downside Risk (DR), and Win-Tie Rate (WTR). All metrics are computed per document and averaged over the evaluation set.

\subsection{Experimental Environment and Setting}
\label{sec:env}

We follow the simulated generative engine setup and evaluation protocol in GEO \citep{aggarwal2024geo} for reproducibility.
All experiments are conducted under a single target generative engine, GPT-4o-mini, using the same decoding parameters as GEO to ensure comparability.

For IF-GEO, all internal calls use the same model.
Unless otherwise specified, we use: query expansion size $N_q{=}5$, suggestions per query $N_s{=}5$, internal temperature $0.2$, and global priority filtering threshold $\tau{=}0.7$.
We analyze the impact of query expansion size in \S\ref{sec:n_analysis}.

Due to the cost of generative-engine evaluation, different experiments use different numbers of queries.
Unless otherwise specified, main results are evaluated on 1{,}000 queries, matching the original GEO test set.
Ablation studies use 250 queries, and the query expansion analysis in \S\ref{sec:n_analysis} uses 500 queries.
All queries are sampled from the same distribution.

\section{Experiments}
\label{sec:experiments}

\begin{table*}[t]
\centering
\caption{Comparison of IF-GEO with baseline methods in robustness and stability across queries. We report variance (VAR$\downarrow$), worst-case performance (WCP$\uparrow$), win--tie rate (WTR$\uparrow$), and downside risk (DR$\downarrow$) for both the primary objective metric (\textit{Obj. Overall}) and subjective metric (\textit{Subj. Average}).}
\label{tab:robustness_main}

\begingroup
\footnotesize
\renewcommand{\arraystretch}{1}
\begin{adjustbox}{max width=\textwidth}
\begin{tabular}{l|cccc|cccc}
\toprule
\multirow{2}{*}{\textbf{Method}} 
& \multicolumn{4}{c|}{\textbf{Objective (Overall)}} 
& \multicolumn{4}{c}{\textbf{Subjective (Average)}} \\
\cmidrule(lr){2-5} \cmidrule(lr){6-9}
 & VAR ($\downarrow$) & WCP ($\uparrow$) & WTR ($\uparrow$) & DR ($\downarrow$)
 & VAR ($\downarrow$) & WCP ($\uparrow$) & WTR ($\uparrow$) & DR ($\downarrow$) \\
\midrule
Tran. SEO    & 0.0147 & -0.0878 & 68.93\% & 0.0062 & 0.0199 & -0.1026 & 88.19\% & 0.0080 \\
Uniq. Word   & 0.0157 & -0.1316 & 61.04\% & 0.0095 & 0.0226 & -0.1361 & 85.34\% & 0.0130 \\
Simp. Expr.  & \underline{0.0116} & -0.1051 & 66.43\% & 0.0058 & \underline{0.0156} & -0.0939 & 88.17\% & 0.0073 \\
Auth. Expr.  & 0.0136 & -0.1018 & 65.84\% & 0.0065 & 0.0200 & -0.1148 & 86.89\% & 0.0089 \\
Flue. Expr.  & 0.0130 & -0.1028 & 68.93\% & 0.0061 & 0.0187 & -0.1084 & 87.38\% & 0.0091 \\
Term. Addi.  & 0.0143 & -0.1113 & 67.31\% & 0.0062 & 0.0216 & -0.1266 & 86.61\% & 0.0103 \\
Cite Sources & 0.0165 & -0.0785 & 72.06\% & 0.0044 & 0.0209 & -0.0892 & \underline{88.93\%} & 0.0070 \\
Quot. Addi.  & 0.0173 & -0.0831 & 70.56\% & 0.0055 & 0.0218 & -0.1013 & 88.25\% & 0.0087 \\
Stat. Addi.  & 0.0173 & -0.0901 & 72.58\% & 0.0060 & 0.0203 & -0.0946 & 88.48\% & 0.0074 \\
RAID         & 0.0166 & -0.1141 & 64.50\% & 0.0088 & 0.0195 & -0.1100 & 82.37\% & 0.0089 \\
Auto-GEO     & 0.0159 & \underline{-0.0511} & \underline{73.56\%} & \underline{0.0043}
             & 0.0203 & \underline{-0.0798} & 84.19\% & \underline{0.0064} \\
\midrule
\textbf{IF-GEO} 
& \textbf{0.0189} & \textbf{-0.0090} & \textbf{80.50\%} & \textbf{0.0023}
& \textbf{0.0116} & \textbf{-0.0419} & \textbf{85.56\%} & \textbf{0.0036} \\
\bottomrule
\end{tabular}
\end{adjustbox}
\endgroup

\end{table*}

\subsection{Main Results}
\label{sec:main_results}

\paragraph{Overall effectiveness.}
Table~\ref{tab:main_results_updated} compares IF-GEO with a diverse set of baselines.
IF-GEO achieves the best performance across all objective and subjective dimensions, reaching an objective overall score of $11.03$ and a subjective average of $5.87$.
Notably, Auto-GEO forms the strongest baseline tier across nearly all dimensions, reflecting the effectiveness of explicitly extracted generative-engine preference rules in shaping citation- and visibility-oriented content. RAID yields limited gains in our simulation. This performance lag likely stems from its singular query trajectory—converging on one refined query from a single initial estimate —which fails to sufficiently span the latent queries or capture heterogeneous user expectations.
Within GEO heuristics, evidence-driven strategies (e.g., \textsc{Cite Sources}, \textsc{Quote Addition}) are consistently strongest, while lexical interventions (e.g., \textsc{Uniq. Word}) can even hurt performance.

\paragraph{Robustness and stability.}
Beyond average gains, Table~\ref{tab:robustness_main} reports stability summaries over the latent queries.
IF-GEO achieves the best objective WCP and the lowest DR, indicating strong protection against query-specific regressions. Notably, a higher VAR does not necessarily imply worse tail stability, as variance captures both positive and negative dispersion. DR and WCP more directly characterize downside behavior across queries.
While Auto-GEO is the strongest baseline on several tail-oriented criteria, it still exhibits larger worst query drops than IF-GEO, and heuristic methods remain more brittle.

Overall, these results confirm that explicitly coordinating revisions to resolve conflicts yields higher and safer holistic visibility than applying isolated, query-agnostic edits.

\subsection{Ablation Study}
\label{sec:ablation}

To quantify each component’s contribution, we ablate IF-GEO by removing conflict resolution, blueprint construction, or the entire instruction fusion stage (Phase~II). Table~\ref{tab:ablation} reveals a clear division of labor. Conflict resolution acts as a safety guardrail: without it, incompatible query-specific edits propagate into execution, producing the largest overall drop and the worst tail behavior (more negative WCP and higher DR). Instruction fusion is the main stabilizer across queries: removing it most strongly reduces reliability (WTR) and amplifies downside risk, consistent with uncoordinated local requests being redundant or misaligned when aggregated. In contrast, blueprint construction primarily affects gain realizability---removing it lowers the achievable overall score while leaving DR essentially unchanged. Together, these results suggest a separation of concerns: conflict resolution and instruction fusion primarily affect stability across queries, while blueprint construction mainly improves the executability of the aggregated edits.

\begin{figure*}[t]
  \centering
  \includegraphics[width=\textwidth]{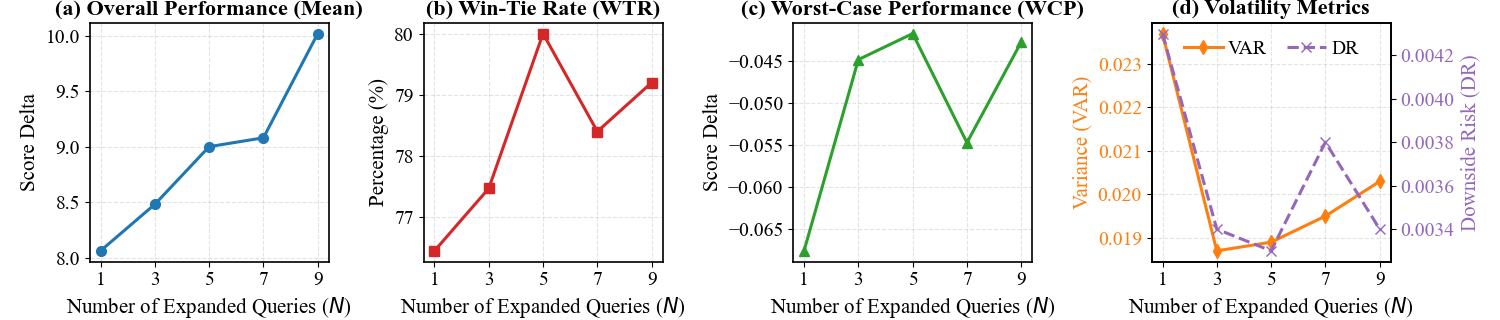}
  \caption{Effect of query expansion size $N$ on performance and stability.
  (a) Overall objective performance (Mean).
  (b) Win--tie rate (WTR).
  (c) Worst-case performance (WCP).
  (d) Volatility metrics: variance (VAR) and downside risk (DR; lower is better).}
  \label{fig:n_analysis}
\end{figure*}

\begin{table}[t]
  \centering
  \caption{Ablation study of IF-GEO. We report the overall objective score (Mean) and stability metrics (VAR, WCP, WTR, DR) to demonstrate the contribution of each component.}
  \label{tab:ablation}
  \resizebox{\columnwidth}{!}{
      \begin{tabular}{l | c | c c c c}
        \toprule
        \textbf{Variant} & \textbf{Mean ($\uparrow$)} & \textbf{VAR ($\downarrow$)} & \textbf{WCP ($\uparrow$)} & \textbf{WTR ($\uparrow$)} & \textbf{DR ($\downarrow$)} \\
        \midrule
        \textbf{IF-GEO (Full)} & \textbf{9.24} & \textbf{0.0156} & \textbf{-0.0328} & 80.80\% & \textbf{0.0021} \\
        \midrule
        w/o Blueprint      & 8.18 & 0.0167 & -0.0517 & \textbf{81.20\%} & \textbf{0.0021} \\
        w/o Instru. Fusion  & 7.07 & 0.0156 & -0.0569 & 74.80\% & 0.0043 \\
        w/o Conflict Res.  & 6.14 & 0.0174 & -0.0713 & 77.20\% & 0.0032 \\
        \bottomrule
      \end{tabular}
  }
\end{table}

\subsection{Impact of query expansion Size}
\label{sec:n_analysis}

We examine the scaling behavior of IF-GEO by varying the number of expanded queries $N\in\{1,3,5,7,9\}$ while keeping other hyperparameters fixed. Figure~\ref{fig:n_analysis} shows that increasing $N$ yields consistent gains: the objective mean performance rises monotonically from $8.06$ at $N{=}1$ to $10.02$ at $N{=}9$ (Figure~\ref{fig:n_analysis}a). Stability also improves overall as query coverage expands, with lower downside risk and less negative worst-case performance. However, these stability gains exhibit diminishing returns beyond $N{=}5$ (Figure~\ref{fig:n_analysis}b--d): WTR peaks at $N{=}5$ (80.00\%) and then fluctuates, while DR and WCP change only marginally from $N{=}5$ to $N{=}9$. Since computation scales roughly linearly with $N$ due to query generation and fusion overhead, we adopt $N{=}5$ as a balanced default that achieves near-peak stability with competitive gains at substantially lower latency.

\subsection{Adaptability Analysis}
\label{sec:adaptability}

We further examine the adaptability of IF-GEO under two realistic variations: changing the underlying generative engine and varying the initial ranking of optimized documents.

\paragraph{Cross-model generalization.}
We evaluate IF-GEO on a different generation model, \textit{Gemini-2.0-Flash}, without any method-specific tuning.
As shown in Appendix~\ref{app:gemini}, IF-GEO consistently achieves the strongest objective gains and favorable stability profiles compared with all baselines.
Notably, while Auto-GEO remains a competitive baseline by leveraging engine-specific preference rules, IF-GEO preserves its advantages in worst-case performance and win--tie rate, suggesting that explicit cross-query consolidation generalizes beyond a single generative engine.

\paragraph{Sensitivity to initial document ranking.}
We further analyze IF-GEO by stratifying documents according to their initial ranking positions prior to optimization.
Results in Appendix~\ref{app:rank} show that IF-GEO yields consistent improvements across all rank buckets, rather than concentrating gains on already well-ranked documents.
In particular, both objective and subjective metrics indicate stable worst-case performance and low downside risk even for lower-ranked inputs, suggesting that IF-GEO improves content robustness rather than merely exploiting positional advantages.

\section{Conclusion}
\label{sec:conclusion}

We study Generative Engine Optimization, aiming to improve a document's visibility across a representative query set while maintaining cross-query stability. We propose \textbf{IF-GEO}, a ``diverge-then-converge'' framework that discovers latent queries, elicits query-specific edit requests (instruction candidates), and fuses them via prioritization, deduplication, and conflict resolution into a unified, executable revision blueprint for blueprint-guided editing. Experiments on GEO-Bench under GPT-4o-mini show that IF-GEO surpasses strong baselines, achieving higher overall gains with lower downside risk. These results underscore the importance of explicitly coordinating competing revision signals for robust GEO beyond single-query tuning.

\section{Limitations}
\label{sec:limitations}

Despite its effectiveness, our work has three main limitations.
First, \textbf{Inference Cost}. The multi-stage ``diverge-then-converge'' workflow involves multiple LLM calls, resulting in higher token consumption than single-pass baselines. 
Second, \textbf{Simulation Gap}. Following standard GEO protocols, we rely on LLM-simulated environments (e.g., GPT-4o-mini). While reproducible, these simulations may not perfectly mirror the commercial engines.
Third, \textbf{Dependency on Query Discovery}. The quality of the Global Revision Blueprint hinges on the representativeness of the mined queries. If the initial query expansion fails to capture the true latent search intent distribution, the subsequent optimization may be misaligned.

\bibliography{custom}

\appendix
\section{Multi-query Competition Diagnostic}
\label{sec:appendix_competition}

This appendix reports a diagnostic experiment that empirically examines the competitive effects induced by \emph{per-query tuning} in Multi-query GEO.
The purpose is not to compare IF-GEO with baselines, but to provide quantitative evidence for the motivation that optimizing a document for one query can concentrate gains on that query and yield uneven outcomes across the remaining queries.

\paragraph{Experimental Setup.}
We use the same dataset, query sets, and evaluation pipeline as in the main experiments.
Each instance contains one document and a representative query set of fixed size $K{=}5$ (constructed following the GEO-Bench setting), with no additional filtering or re-sampling.

For each document, we elicit query-conditioned edit requests independently for all queries using the same request format as IF-GEO.
We then uniformly sample one query as the target $i^*$ and perform a single per-query optimization pass that rewrites the document using only the requests associated with $i^*$.
No information from the other queries is used during rewriting, and no cross-query fusion is performed.

We evaluate visibility before and after per-query tuning under all $K$ queries using the primary objective visibility metric (Obj.\ Overall), following the definitions in Section~\ref{sec:metrics}.

\paragraph{Metrics.}
We report three statistics for each category: (i) \emph{Mean Gain}, the average visibility improvement; (ii) $P(\text{gain}<0)$, the rate of negative gain; and (iii) \emph{Downside Magnitude (DM)}, defined as $\mathbb{E}[-\min(0,\text{gain})]$, which measures the expected magnitude of negative outcomes.
DM is an interpretable \emph{L1} analogue of the main-text downside risk metric DR (which squares negative gains), and is reported here only for diagnostic analysis.
For relative spillover, we compute the same statistics over $r_j$.

\paragraph{Results.}
Table~\ref{tab:competition_diagnostic} summarizes the outcomes over 202 records (200 documents, 808 $(i^*,j)$ pairs) under the primary objective metric (Obj.\ Overall).
per-query tuning exhibits a clear asymmetric improvement pattern across the query set.
The optimized (target) query $i^*$ achieves a substantially larger mean gain (0.277), with a relatively low negative-gain rate ($P(\text{gain}<0)=0.124$) and small downside magnitude (DM=0.017).
In contrast, the non-target queries ($j\neq i^*$) show weaker and less reliable improvements: while the mean gain remains positive (0.087), the negative-gain rate more than doubles (0.306) and DM increases to 0.036.

Relative spillover further indicates systematic concentration of gains on the target query.
On average, non-target queries lag behind the optimized query by 0.189 (mean spillover $=-0.189$), and in 69.2\% of query pairs the spillover is negative ($P(r_j<0)=0.692$).
Moreover, the large spillover downside magnitude (DM=0.228) suggests that when non-target queries fall behind, the gap is often substantial rather than marginal.

\paragraph{Discussion.}
Overall, this diagnostic indicates that single-query tuning primarily induces a pronounced gain allocation skew within the representative query set.
Rather than causing pervasive absolute degradation on other queries, it tends to concentrate improvements on the optimized query while leaving the remaining queries with smaller and notably less stable benefits.
This pattern motivates the need for \emph{global coordination} over the query set---to achieve reliable and well-distributed improvements under a shared content budget---as pursued by IF-GEO in the main method.

\begin{table}[t]
\centering
\small
\setlength{\tabcolsep}{5.5pt}
\caption{Gain allocation skew under per-query tuning (primary objective metric, Obj.\ Overall).
We report mean gain, the negative-gain rate $P(\text{gain}<0)$, and downside magnitude (DM) for the optimized query ($i^*$), the non-target queries ($j\neq i^*$), and relative spillover ($r_j$), aggregated over all $(i^*,j)$ pairs.}
\label{tab:competition_diagnostic}
\begin{tabular*}{\linewidth}{@{\extracolsep{\fill}}lccc}
\toprule
 & Mean Gain & $P(\text{gain}<0)$ & DM \\
\midrule
Optimized query & 0.277 & 0.124 & 0.017 \\
Non-target queries & 0.087 & 0.306 & 0.036 \\
\midrule
Relative spillover & -0.189 & 0.692 & 0.228 \\
\bottomrule
\end{tabular*}
\end{table}

\section{Prompts Used in IF-GEO}
\label{sec:appendix_prompts}

This section documents the main prompts used in IF-GEO. 
All prompts are shown in their original form to ensure full transparency and reproducibility.
They correspond to different steps of the IF-GEO pipeline and are executed sequentially.

\paragraph{Query Mining.}

\begin{tcolorbox}[ifgeoGray]
\textbf{System Prompt:}

You are a search log analyst. You study how real users search on the web.
Your task is to infer what queries users are most likely to type when they are trying to find or understand the content of a given webpage.

\medskip
\textbf{Task Description:}

Based on the webpage content provided by the user, infer \emph{realistic search queries} that would naturally lead them to this page.

\medskip
\textbf{Your Goals:}
\begin{itemize}[leftmargin=*,nosep]
  \item Generate \textbf{\{num\_queries\} queries} that reflect \textbf{different aspects of the main theme}.
  \item Focus on the central topic, \textbf{not specific details} (e.g., trivial facts).
  \item Ensure queries are \textbf{natural}, \textbf{diverse}, and \textbf{not simple paraphrases}.
  \item For each query, estimate a \textbf{probability score (0--100)} reflecting real-world likelihood.
\end{itemize}

\medskip
\textbf{Output Format:}

You must return ONLY a valid JSON object. Do not include markdown formatting or any other text.
The format must be exactly:
\begin{verbatim}
{
  "queries": [
    {
      "query": "the inferred user query string",
      "probability": 85
    }
  ]
}
\end{verbatim}
\end{tcolorbox}

\paragraph{Query-specific Request Generation.}

\begin{tcolorbox}[ifgeoGray]
\textbf{System Prompt:}

You are an assistant who reviews how well a given webpage can answer a specific user query.
Your goal is to identify concrete weaknesses in the webpage and propose clear, actionable revision suggestions.

\medskip
\textbf{Task Description:}

You will receive a \textbf{user query} and a \textbf{webpage text}.

Your task:
\begin{itemize}[leftmargin=*,nosep]
  \item Evaluate how well the webpage text could answer the query.
  \item Identify \textbf{specific weaknesses} in the text that may prevent a strong answer.
  \item Produce \textbf{up to \{suggestions\_num\} revision suggestions} that would improve the webpage's ability to answer the query.
\end{itemize}

\medskip
\textbf{Requirements for Each Suggestion:}
\begin{itemize}[leftmargin=*,nosep]
  \item \textbf{Target Specificity:} Refer to a \textbf{specific part, sentence, or paragraph} of the webpage.
  \item \textbf{Excerpt:} Include a \textbf{short excerpt} (quoted or partially quoted) to make the target explicit.
  \item \textbf{Actionable:} Describe \textbf{how to modify} that part (e.g., add missing details, clarify wording, restructure, simplify).
  \item \textbf{Necessity Score:} Include a \textbf{necessity score (0--100)} indicating importance for answering the query.
\end{itemize}

\medskip
\textbf{Output Format (JSON Only):}

Return ONLY a valid JSON object. Do not include markdown formatting or any other text.
The format must be exactly:
\begin{verbatim}
{
  "suggestions": [
    {
      "excerpt": "short excerpt",
      "suggestion": "revision details",
      "necessity": 85
    }
  ]
}
\end{verbatim}
\end{tcolorbox}

\paragraph{Prioritization and Deduplication.}

\begin{tcolorbox}[ifgeoGray]
\textbf{System Prompt:}

You are a Senior Editor consolidating feedback from multiple readers.

\medskip
\textbf{Task:}

You are provided with a list of revision suggestions grouped by user queries.
Your goal is to \textbf{flatten} this list and \textbf{merge} duplicate suggestions that target the same content or issue.

\medskip
\textbf{Rules:}
\begin{itemize}[leftmargin=*,nosep]
  \item \textbf{De-duplicate:} Merge suggestions ONLY if they target the \textbf{same semantic topic} AND apply to the \textbf{same approximate location (excerpt)} in the text.
  \begin{itemize}[leftmargin=*,nosep]
    \item \emph{Constraint:} Do NOT merge suggestions that target widely separated paragraphs (e.g., ``Intro'' vs ``Conclusion''), even if they are thematically related. Keep them as separate items with the same topic tag.
  \end{itemize}
  \item \textbf{Filter:} Discard suggestions with \texttt{necessity} $<$ 60 unless they address a critical factual error.
  \item \textbf{Standardize:} Output a flat list. Assign a short \texttt{topic} tag to each.
\end{itemize}

\medskip
\textbf{Output Format (JSON Only):}

Return ONLY a valid JSON list. Do not include markdown formatting.
Example:
\begin{verbatim}
[
  {
    "id": "suggest_1",
    "topic": "eg.Comparison Table",
    "excerpt": "short excerpt",
    "suggestion": "modify-suggestion content",
    "necessity": 95
  }
]
\end{verbatim}
\end{tcolorbox}

\paragraph{Conflict Resolution.}

\begin{tcolorbox}[ifgeoGray]

\textbf{System Prompt:}

You are a Logic Arbiter resolving content revision conflicts.

\medskip
\textbf{Task:}

Examine the provided list of suggestions. Detect logical conflicts (where two or more suggestions cannot be simultaneously implemented) and resolve them according to the following priority-based rules.

\medskip
\textbf{Rules:}
\begin{itemize}[leftmargin=*,nosep]
  \item \textbf{Detect Conflicts:} Look for instructions targeting the same \texttt{excerpt} or topic that are mutually exclusive (e.g., ``Delete section'' vs ``Expand section'').
  \item \textbf{Resolve:}
  \begin{itemize}[leftmargin=*,nosep]
    \item If conflict exists, prioritize the instruction with higher \texttt{necessity}.
    \item If instructions are compatible (e.g., ``Shorten text'' AND ``Add link''), combine them into one instruction.
  \end{itemize}
  \item \textbf{Output:} A final, clean list of instructions.
\end{itemize}

\medskip
\textbf{Output Format (JSON Only):}

Return ONLY a valid JSON list. Do not include markdown formatting.
Example:
\begin{verbatim}
[
  {
    "id": "suggest_1",
    "excerpt": "original excerpt",
    "suggestion": "instruction text..."
  }
]
\end{verbatim}
\end{tcolorbox}

\paragraph{Blueprint Construction.}

\begin{tcolorbox}[ifgeoGray]
\textbf{System Prompt:}

You are a Content Architect \& Strategist.

\medskip
\textbf{Task:}

You are given a webpage and a set of logic-checked revision instructions.
Your goal is to create a \textbf{Master Revision Plan} that maps these instructions to the webpage structure.

\medskip
\textbf{Process:}
\begin{itemize}[leftmargin=*,nosep]
  \item \textbf{Analyze Structure:} Read the webpage to understand its current flow (Intro $\rightarrow$ Body $\rightarrow$ Conclusion).
  \item \textbf{Map \& Group:} Assign each instruction to a specific logical section of the webpage.
  \begin{itemize}[leftmargin=*,nosep]
    \item If multiple instructions target the same section (e.g., ``Add history'' and ``Fix grammar'' both in Intro), \textbf{GROUP} them into one plan item.
    \item If an instruction requires a new section, decide the best insertion point to maintain flow.
  \end{itemize}
  \item \textbf{Plan:} Output a structured blueprint.
\end{itemize}

\medskip
\textbf{Output Format (JSON Only):}

Return ONLY a valid JSON object. Do not include markdown formatting.
Example:
\begin{verbatim}
{
  "revision_blueprint": [
    {
      "section_name": "Introduction",
      "target_location": "...",
      "modification_intent": "...",
      "directives": [
        "Integrate instruction #1: ...",
        "Integrate instruction #5: ..."
      ],
      "format_note": "Keep as text paragraphs."
    }
  ]
}
\end{verbatim}
\end{tcolorbox}

\paragraph{Blueprint-Guided Revision.}

\begin{tcolorbox}[ifgeoGray]
\textbf{System Prompt:}

You are an expert Content Engineer and GEO (Generative Engine Optimization) Specialist.

\medskip
\textbf{Task:}

Your task is to rewrite a given webpage content by strictly following a provided ``Revision Blueprint''.

\medskip
\textbf{Rules:}
\begin{itemize}[leftmargin=*,nosep]
  \item \textbf{Full Output:} You must output the \textbf{ENTIRE} rewritten webpage. Do not summarize, do not omit sections, and do not use placeholders like ``[...rest of text remains same...]''.
  \item \textbf{Strict Execution:} Implement every directive in the Blueprint (e.g., inserting tables, adding links, rewriting paragraphs, adding new sections).
  \item \textbf{Preservation:} For sections NOT mentioned in the Blueprint, preserve the original text and structure exactly as is.
  \item \textbf{Formatting:} Use standard Markdown. If the Blueprint asks for a table, render a valid Markdown table.
  \item \textbf{Tone:} If the Blueprint specifies a tone change (e.g., ``professional''), apply it effectively to the target section.
\end{itemize}

\medskip
\textbf{Output:}

Return only the final rewritten content in Markdown format.
\end{tcolorbox}


\section{Qualitative Case Study}
\label{app:case}

\noindent
This appendix presents one end-to-end example, organized by IF-GEO steps.
For readability, we use ellipses (\dots) to indicate omitted spans.

\subsection{Input Document}
\label{app:caseA-input}

\noindent
We start from the original document snippet, where terminology around
\emph{coagulopathy} is easy to misinterpret without careful disambiguation.

\begin{tcolorbox}[ifgeoGray]
\textbf{Document snippet}\\
``Coagulopathy is a condition in which the blood's ability to for\dots, leading to a tendency towards prolonged or excessive bleeding.\\
It can occur spontaneously or following an injury or medical procedure.\\
\dots\ Coagulopathies are sometimes mistakenly referred to as `clotti\dots haracterized by a predisposition to excessive clot formation.\ \dots''
\end{tcolorbox}

\subsection{Query Mining}
\label{app:caseA-intents}

\noindent
IF-GEO first mines a diverse query set (queries) that could reasonably lead to the document,
with associated weights (here shown as probabilities).

\begin{tcolorbox}[ifgeoGray]
\begin{itemize}[leftmargin=*,nosep]
  \item \texttt{what is coagulopathy and its symptoms} \hfill (\textit{p}=0.90)
  \item \texttt{causes and treatment options for coagulopathy} \hfill (\textit{p}=0.85)
  \item \texttt{how does coagulopathy affect bleeding risk} \hfill (\textit{p}=0.80)
  \item \texttt{difference between coagulopathy and clotting disorders} \hfill (\textit{p}=0.75)
  \item \texttt{emergency management of coagulopathy in trauma patients} \hfill (\textit{p}=0.70)
\end{itemize}
\end{tcolorbox}

\subsection{Query-specific Request Generation.}
\label{app:caseA-suggestions}

\noindent
Given each query, the system generates localized edit requests.
Here we show the merged pool of core suggestions before conflict-aware instruction fusion.

\begin{tcolorbox}[ifgeoGray]
\begin{itemize}[leftmargin=*,nosep]
  \item \textbf{Excerpt:} \emph{Coagulopathy is a condition in which the blood's ability to form clots is impaired\dots}\\
        \textbf{Suggestion:} Simplify this definition to make it more accessible. For example\dots bleeding.' This will help readers grasp the concept more easily.\\
        \textbf{Necessity:} 75

  \item \textbf{Excerpt:} \emph{Coagulopathies are sometimes mistakenly referred\dots as `clotting disorders,' but they are actually the opposite\dots}\\
        \textbf{Suggestion:} Provide a brief explanation or definition of `clotting disorders'\dots coagulopathy is the opposite of what they may mistakenly believe.\\
        \textbf{Necessity:} 70

  \item \textbf{Excerpt:} \emph{Coagulopathies are sometimes mistakenly referred\dots, characterized by a predisposition to excessive clot formation.}\\
        \textbf{Suggestion:} Clarify the distinction between coagulopathy and clotting disord\dots rs understand why coagulopathy leads to increased bleeding risk.\\
        \textbf{Necessity:} 80

  \item \textbf{Excerpt:} \emph{Coagulopathies are sometimes mistakenly referred to as ``clotti\dots, characterized by a predisposition to excessive clot formation.}\\
        \textbf{Suggestion:} Clarify the distinction between coagulopathy and clotting disord\dots te clot formation and explain how they differ from coagulopathy.\\
        \textbf{Necessity:} 90
\end{itemize}
\end{tcolorbox}

\subsection{Conflict-Aware Instruction Fusion}
\label{app:caseA-blueprint}

\noindent
Multiple suggestions target overlapping anchors but imply incompatible framing.
IF-GEO resolves these conflicts by producing a compact, globally consistent revision blueprint.

\begin{tcolorbox}[ifgeoGray]
\textbf{Section name:} Definition and Terminology Clarification\\
\textbf{Target location:} The opening definition sentence starting with \dots\ followed by the sentence mentioning `clotting disorders' \dots\\
\textbf{Modification intent:} Make the lead definition more accessible w\dots\ biguation and unifying terminology to prevent misinterpretation.\\[2pt]

\textbf{Directives:}
\begin{enumerate}[leftmargin=*,nosep]
  \item Integrate instruction \#1: Rewrite the opening definitio\dots\ ding multi-term contrasts inside this first definition sentence.
  \item Integrate instruction \#2: Add a compact disambiguation (1--2 se\dots\ aired clot formation and bleeding risk, not excessive clotting).
  \item Integrate instruction \#3: Resolve inconsistent wording by unif\dots\ reader cannot misinterpret coagulopathy as ``excessive clotting''.
\end{enumerate}

\textbf{Format note:} Keep within the opening paragraph; add at most 1--2 clarification sentences and do not introduce new sections.
\end{tcolorbox}

\subsection{Blueprint-Guided Revision}
\label{app:caseA-after}

\noindent
Finally, the document is edited under the blueprint constraints to balance readability
and terminological correctness with minimal, localized changes.

\begin{tcolorbox}[ifgeoGray]
\textbf{Document snippet (after, expected).}\\
``Coagulopathy is a condition in which the blood's ability to for\dots m clots is impaired, leading to prolonged or excessive bleeding.\\
It can occur spontaneously or following an injury or medical procedure.\\
\dots\ The term is sometimes confused with `clotting disorders,' a br\dots n increased risk of bleeding, rather than excessive clotting.\ \dots''
\end{tcolorbox}


\subsection{Token Breakdown of IF-GEO}
\label{app:tokens_ifgeo}

Unlike conventional single-pass GEO baselines that rewrite a document once,
IF-GEO performs a two-stage \emph{``diverge-then-converge''} workflow,
where each stage targets a distinct task.
Table~\ref{tab:token_ifgeo} reports the average token consumption per document for each step,
aggregated over the evaluation set (prompt+completion tokens).

\begin{table}[t]
\centering
\small
\setlength{\tabcolsep}{6pt}
\caption{Average token consumption per stage in IF-GEO (per document).}
\label{tab:token_ifgeo}
\begin{tabular}{l|c}
\toprule
\textbf{Stage} & \textbf{Avg. Tokens} \\
\midrule
Query Mining & 1270.6 \\
Edit Request Generation & 1749.8 \\
Instruction Fusion & 4487.6 \\
Blueprint-Guided Revision & 2819.8 \\
\midrule
\textbf{Total} & \textbf{10327.7} \\
\bottomrule
\end{tabular}
\end{table}

The dominant cost comes from \textbf{Instruction Fusion}, which consolidates query-specific edit requests into a single executable \emph{global revision blueprint}.
This stage involves joint reasoning over partially overlapping directives (e.g., deduplication, prioritization, and arbitration),
and therefore scales with the diversity of intents and the number of candidate requests rather than redundant text generation.

\subsection{Comparison with Single-Pass GEO Baselines}
\label{app:tokens_baselines}

For reference, Table~\ref{tab:token_baselines} reports the token consumption of representative single-pass GEO baselines,
each executed once per document under the same model configuration.
These baselines apply localized rewriting heuristics without explicitly coordinating across the representative query set.

\begin{table}[t]
\centering
\small
\setlength{\tabcolsep}{6pt}
\caption{Average token consumption of single-pass GEO baselines (per document).}
\label{tab:token_baselines}
\begin{tabular}{l|c}
\toprule
\textbf{Method} & \textbf{Avg. Tokens} \\
\midrule
Tran.\ SEO     & 2653.9 \\
Uniq.\ Word    & 2282.8 \\
Simp.\ Expr.   & 2204.3 \\
Auth.\ Expr.   & 2483.7 \\
Flue.\ Expr.   & 2171.6 \\
Term.\ Addi.   & 2392.4 \\
Cite Sources   & 2535.0 \\
Quot.\ Addi.   & 2589.5 \\
Stat.\ Addi.   & 2802.2 \\
\bottomrule
\end{tabular}
\end{table}

As expected, single-pass baselines incur substantially lower token usage,
since they do not explore multiple queries nor perform cross-query consolidation.

\begin{table}[t]
\centering
\scriptsize
\setlength{\tabcolsep}{3.5pt}
\renewcommand{\arraystretch}{0.95}
\caption{Objective visibility on Gemini-2.0-Flash (primary objective metric: \textit{Obj.\ Overall}).
We report mean improvement (Mean$\uparrow$) and stability summaries (VAR$\downarrow$, WCP$\uparrow$, WTR$\uparrow$, DR$\downarrow$).}
\label{tab:gemini_objective}
\begin{tabular}{lccccc}
\toprule
Method & Mean & VAR & WCP & WTR & DR \\
\midrule
Tran.\ SEO   &  1.93 & 0.0252 & -0.1313 & 70.28\% & 0.0089 \\
Uniq.\ Word  & -5.99 & 0.0290 & -0.2187 & 56.12\% & 0.0279 \\
Simp.\ Expr. & -0.88 & 0.0196 & -0.1496 & 65.24\% & 0.0127 \\
Auth.\ Expr. & -0.02 & 0.0241 & -0.1474 & 65.08\% & 0.0137 \\
Flue.\ Expr. & -1.93 & 0.0267 & -0.1834 & 62.56\% & 0.0176 \\
Term.\ Addi. &  1.31 & 0.0231 & -0.1354 & 69.33\% & 0.0102 \\
Cite Sources &  2.54 & 0.0246 & -0.1156 & 74.21\% & 0.0089 \\
Quot.\ Addi. &  3.10 & 0.0321 & -0.1284 & 72.58\% & 0.0099 \\
Stat.\ Addi. &  0.25 & 0.0190 & -0.1345 & 71.57\% & 0.0108 \\
RAID         &  2.80 & 0.0240 & -0.1248 & 71.43\% & 0.0099 \\
Auto-GEO     & 12.99 & 0.0416 & -0.0578 & 78.91\% & 0.0083 \\
\midrule
\textbf{IF-GEO} & \textbf{14.17} & \textbf{0.0386} & \textbf{-0.0435} & \textbf{84.07\%} & \textbf{0.0054} \\
\bottomrule
\end{tabular}
\end{table}

\section{Performance Stratified by Initial Document Ranking}
\label{app:rank}

This appendix complements the \emph{sensitivity to initial document ranking} analysis in Section~\ref{sec:adaptability}.
We stratify evaluation instances by their \emph{initial ranking position} prior to optimization and report rank-bucketed performance of IF-GEO.
The goal is to verify that improvements are not confined to already well-ranked inputs, but persist across ranking strata.

\paragraph{Rank buckets.}
Each instance is assigned to one of five buckets (\textbf{Rank1}--\textbf{Rank5}) according to its initial ranking position under the same evaluation setting as the main experiments, where \textbf{Rank1} corresponds to higher-ranked inputs and \textbf{Rank5} to lower-ranked inputs.
All instances and metrics follow the main evaluation protocol; only the reporting is stratified by rank.

\paragraph{Metrics.}
For each bucket, we report mean improvement (Mean) and cross-query stability summaries (VAR, WCP, WTR, DR) for both the primary objective metric (\textit{Obj.\ Overall}) and the subjective metric (\textit{Subj.\ Average}), using the same definitions as in Section~\ref{sec:metrics}.

\begin{table*}[t]
\centering
\caption{Rank-stratified performance of IF-GEO by initial ranking position prior to optimization.
We report mean improvement (Mean$\uparrow$) and stability summaries (VAR$\downarrow$, WCP$\uparrow$, WTR$\uparrow$, DR$\downarrow$) for the primary objective metric (\textit{Obj.\ Overall}) and the subjective metric (\textit{Subj.\ Average}).}
\label{tab:rank_stratified_main}
\resizebox{\textwidth}{!}{%
\begin{tabular}{l|ccccc|ccccc}
\toprule
\multirow{2}{*}{\textbf{Rank}} 
& \multicolumn{5}{c|}{\textbf{Obj.\ Overall}} 
& \multicolumn{5}{c}{\textbf{Subj.\ Average}} \\
\cmidrule(lr){2-6} \cmidrule(lr){7-11}
& Mean ($\uparrow$) & VAR ($\downarrow$) & WCP ($\uparrow$) & WTR ($\uparrow$) & DR ($\downarrow$)
& Mean ($\uparrow$) & VAR ($\downarrow$) & WCP ($\uparrow$) & WTR ($\uparrow$) & DR ($\downarrow$) \\
\midrule

IF-GEO
& 11.03 & 0.0156 & -0.0090 & 80.50\% & 0.0023
& 5.87 & 0.0116 & -0.0419 & 85.56\% & 0.0036 \\
\midrule

Rank1 
& 13.49 & 0.0174 & 0.0084  & 77.92\% & 0.0033
& 6.39  & 0.0140 & -0.0431 & 85.30\% & 0.0054 \\

Rank2 
& 8.56  & 0.0121 & -0.0281 & 77.36\% & 0.0031
& 6.78  & 0.0108 & -0.0292 & 84.64\% & 0.0038 \\

Rank3 
& 8.71  & 0.0159 & -0.0209 & 82.76\% & 0.0007
& 5.65  & 0.0092 & -0.0223 & 87.59\% & 0.0008 \\

Rank4 
& 12.24 & 0.0140 & 0.0196  & 87.14\% & 0.0006
& 5.20  & 0.0119 & -0.0592 & 87.55\% & 0.0028 \\

Rank5 
& 12.14 & 0.0189 & -0.0154 & 81.43\% & 0.0022
& 4.72  & 0.0114 & -0.0587 & 84.29\% & 0.0038 \\
\bottomrule
\end{tabular}%
}
\end{table*}

\paragraph{Discussion.}
IF-GEO yields consistent improvements across all rank buckets for both \textit{Obj.\ Overall} and \textit{Subj.\ Average}, indicating that gains are not concentrated exclusively on higher-ranked inputs.
In particular, lower-ranked buckets (Rank4--Rank5) still achieve sizable mean improvements while maintaining favorable tail-oriented stability (e.g., competitive WCP/WTR and low DR), supporting the claim in Section~\ref{sec:adaptability} that IF-GEO improves content robustness rather than merely exploiting positional advantages.

\section{Evaluation on Gemini-2.0-Flash}
\label{app:gemini}

This appendix complements the \emph{cross-model generalization} analysis in Section~\ref{sec:adaptability}.
We replace the underlying generative engine with \textit{Gemini-2.0-Flash} and re-evaluate all methods under the same dataset, query sets, and evaluation protocol as in the main experiments.
Due to interface limitations, we report only the primary objective visibility metric (\textit{Obj.\ Overall}) in this setting.

\paragraph{Setup.}
All GEO baselines, RAID, Auto-GEO, and IF-GEO are re-run on Gemini-2.0-Flash without any method-specific tuning.
All configurations follow the main evaluation pipeline; the only change is the underlying generative engine.

\paragraph{Metrics.}
We report mean improvement (Mean) and cross-query stability summaries (VAR, WCP, WTR, DR) computed over the representative query set, following the same definitions as in Section~\ref{sec:metrics}.


\paragraph{Discussion.}
Table~\ref{tab:gemini_objective} shows that IF-GEO remains strong under a different generative engine, achieving the largest mean improvement and favorable tail-oriented stability (e.g., higher WTR and lower DR) among all methods.
The overall trend is consistent with the main results, supporting the cross-model generalization claim in Section~\ref{sec:adaptability}.

\end{document}